# Structure and energetics of hydrogen chemisorbed on a single graphene layer to produce graphane

Abdiravuf A. Dzhurakhalov [a,b,*], Francois M. Peeters [a]

[a] *Department of Physics, University of Antwerp, Groenenborgerlaan 171, B-2020 Antwerpen, Belgium*

[b] *Department of Mathematics and Computer Science, University of Antwerp, Middelheimlaan 1, B-2020 Antwerpen, Belgium*

**Abstract**

Chemisorption of hydrogen on graphene is studied using atomistic simulations with the second generation of reactive empirical bond order Brenner inter-atomic potential. The lowest energy adsorption sites and the most important metastable sites are determined. The H concentration is varied from a single H atom, to clusters of H atoms up to full coverage. We found that when two or more H atoms are present, the most stable configurations of H chemisorption on a single graphene layer are *ortho* hydrogen pairs adsorbed on one side or on both sides of the graphene sheet. The latter has the highest hydrogen binding energy. The next stable configuration is the *ortho-para* pair combination, and then *para* hydrogen pairs. The structural changes of graphene caused by chemisorbed hydrogen are discussed and are compared with existing experimental data and other theoretical calculations. The obtained results will be useful for nanoengineering of graphene by hydrogenation and for hydrogen storage.

---

[*] Corresponding author. Fax: +32 3 2653204.

E-mail: Abdiravuf.Dzhurakhalov@ua.ac.be

## 1. Introduction

In earlier works graphene was used as a model system to describe the properties of carbon nanotubes, graphite nanofibres and fullerenes. The recent discovery of stable single graphene layers [1] makes it a very interesting study object for novel properties of two dimensional electrons. Their properties can be changed by the chemisorption of atoms or molecules. Of particular interest is the opening of an energy gap in the electron spectrum which will allow turning off and on the electrical conduction in a graphene field effect transistor. The hydrogenation of carbon-based materials has gained renewed theoretical and experimental interest because of its importance for hydrogen storage [2-9]. In previous works, it was shown that hydrogen chemisorption corrugates the graphene sheet (in fullerene, carbon nanotube [10], graphite [11] and graphene [12]), and transforms it from a semimetal into a semiconductor [13] and can even induce magnetic moments [14-16]. As demonstrated in [17], hydrogenation of graphene ribbons can be used as a unique method of synthesizing carbon nanotubes (CNT) with controllable diameter and chirality. Recently it was shown that conducting and semiconducting, magnetic and nonmagnetic graphene nanoroads [18] as well as graphene quantum dots [19] with regular edges can be patterned on a graphene sheet by selective hydrogenation. Advantage of this method over pure graphene nanoribbons or dots is that various tunable patterns with desirable properties can be made on the same graphene sheet [18-20].

A unique feature of a single graphene layer is that it has identical surfaces on both sides in contrast to fullerene, carbon nanotube and graphite. Experimental studies show that hydrogenation proceeds at a higher rate for a single graphene layer and that the net H atom sticking probability at 300K exceeds by a factor of 15 that for double layers, demonstrating the enhanced chemical reactivity of a single sheet of graphene [21]. A density functional (DFT) calculation of H atom adsorption positioned at several highly symmetric sites of graphene with rigid and relaxed carbon atoms shows that the largest binding energy is found when the H atom is on top of carbon and it leads to a shift of the nearest C atom by 0.33 Å out of the plane of the graphene layer [22]. The same site was found as the most stable one in [23]. But at the same time it was found that the migration energy barrier for a hydrogen atom is very low resulting in instability of all sites at room temperature. Recently, it was shown that a single H atom adsorbed on a top site of graphene can be distinguished by DFT and STM image calculations from its closely-spaced (*ortho* hydrogen) pairs [24]. Both experiment [25] and calculations [5,24,26,27] found that this *ortho* hydrogen pair is energetically the most stable configuration as compared to *para*, *meta* and other hydrogen pairs.

The curvature, buckling and tensile stress of a single graphene layer make it more chemical active to hydrogen adsorption. Even a single H atom results in a local buckling of the graphene layer [10,11,28]. These structural distortions have been studied in detail for the case of single and two hydrogen atoms using the DFT technique [16], and the upward shift of the C atom beneath H was found to be ~0.3 Å in the case of a single H and ~0.35-0.64 Å for a large H concentration in the case of H pairs. In addition, it was found that by using DFT calculations [28] the size and shape of the substrate can affect the adsorption energy and consequently the shape of the lowest energy hydrogen pair.

To our knowledge there are only few works that discuss hydrogen adsorption on graphene when more than two H atoms are present [5,13,25,27,29,30]. The recently experimentally observed starlike shape of STM images [25] was interpreted as due to the chemisorption of three hydrogen atoms [25,27]. However, the suggested geometrical configuration [27] may be specific for that particular experimental condition where the sample was annealed at 570 K and it is also known that the substrate (i.e. its nanoscale topography) may induce local stress in the graphene layer.

In spite of several studies on the hydrogenation of carbon-based materials a systematic investigation of both the energetics and the structural changes of a single graphene layer and the stable configurations of chemisorbed hydrogen atoms is still lacking. In this paper, we study the interaction of H atoms with a single graphene layer using the energy minimization method, and discuss the structural changes of graphene caused by hydrogen. The purpose of this investigation is to obtain the complete morphology of these structures versus the number of hydrogen atoms within one approach. We will compare systematically the configurations, bond lengths and hydrogen binding energies of hydrogenated graphene as predicted by our calculation with available theoretical and experimental data.

**2. Simulation method**

The different structures of a partially and full hydrogenated single graphene layer are studied by the energy minimization method where we use the conjugate gradient algorithm. As energy is a function of the different degrees of freedom, i.e. bond lengths, bond angles, and dihedrals, this method finds the energetically preferred conformations of the system, which is equivalent in locating all the minima of its energy function. Both the lowest energy state and the most important other local minima states are studied in the present paper.

The interatomic potentials of carbon based materials were calculated by *ab initio* and empiric methods. In general, *ab initio* methods are considered to be more reliable and accurate, but, in practice, they are very time-consuming and limited to relative small unit cells [31]. On the other hand, the empiric or parametric methods give the interaction potential in an analytical form and are bassed on fitting parameters determined from a comparison with experimental data. Due to the analytical form of the potential it is convenient for subsequent calculations of other properties of the material. Here we use the recently modified version of the Brenner potential which was parameterized specifically for carbon and hydrocarbon systems. The expression of this potential and its parameters used in our calculation can be found in [32].

As was shown in [33-36] this potential describes very well the carbon-carbon interaction. Here we show additionally its fitness in describing hydrocarbon systems. In Table 1 the binding energy, the bond length and the bond angles of some hydrocarbons as calculated by the modified Brenner potential which agree very well with available experimental and other theoretical data. This validates our model which we will apply for graphene and its interaction with atomic H. Additionally, calculations carried out by the *ab-initio* method and by the Brenner potential gave the same results in synthesizing CNT using hydrogenation of graphene [17]. This indicates also the correctness of the Brenner potential for study of hydrogen adsorption on graphene.

The initial configuration of the graphene layer was chosen as a planar honeycomb structure with various lattice constants. Starting from an arbitrary initial configuration the final configuration is found by minimizing the energy of the system. If the energy and the configuration of the obtained minimum are different from the previous ones, this configuration is considered as new. Preliminary test on different sizes of simulation unit cell consisting of 32, 112, 448, 1792, 7168 and 28672 atoms shows that a unit cell consisting of 112 carbon atoms is sufficient for the simulation of hydrogenation, i.e. the results are the same for larger cells. That is why further on we use 112-atom single graphene layer with periodic boundary conditions applied in the lateral plane. For a single H atom several high symmetric adsorption sites (with the exception of the center of the benzene ring, because for this site the C-H distance is too large to lead to a chemical bond) were studied. For the H pairs, only on top and bridge sites were investigated while for the other cases only top sites were considered.

Table 1. The binding energy, bond length and angle for some hydrocarbons as obtained within the present approach (**bold**), by the tight-binding method [37] and from experimental data[a,b] [38-40].

| Hydro-carbons | Binding energy (eV/atom) | Bond length (Å) | | Angle ($^0$) | |
|---|---|---|---|---|---|
| | | C-C | C-H | C-C-H | H-C-H |
| $C_2H_2$ 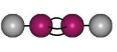 | **4.39**<br>4.54[37]<br>4.29[a,b] | **1.206**<br>1.226[37]<br>1.203[b]<br>1.205[38] | **1.09**<br>1.079[37]<br>1.063[b]<br>1.06[38] | **180**<br><br><br>180[38] | |
| $C_2H_4$ 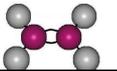 | **4.07**<br>3.96[37]<br>3.94[a,b] | **1.318**<br>1.327[37]<br>1.339[b]<br>1.314[39] | **1.09**<br>1.097[37]<br>1.086[b]<br>1.071[39] | **124.4**<br><br><br>121.2[39] | **111.2**<br><br><br>117.6[39] |
| $CH_4$ 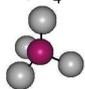 | **3.64**<br>3.40[37]<br>3.51[a,b] | | **1.09**<br>1.10[37]<br>1.094[b]<br>1.094[40] | | **109.5**<br><br><br>109.5[40] |
| $C_6H_6$ 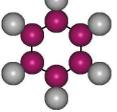 | **4.94**<br>4.82[37]<br>4.79[a,b] | **1.385**<br>1.407[37]<br>1.397[b] | **1.09**<br>1.095[37]<br>1.084[b] | | |

[a] http://webbook.nist.gov/chemistry/
[b] http://srdata.nist.gov/cccbdb/

## 3. Results and discussion

Here we present the energetics and structural changes in the graphene sheet caused by the adsorption of different concentrations of hydrogen atoms from pure graphene to its fully hydrogenated counterpart – graphane. First an infinite graphene sheet with a few hydrogen atoms is studied for different adsorption positions. Then hydrogen linear structures bound on the graphene sheet as well as half and fully saturated graphane structures are studied.

*3.1 Initial hydrogenation stages of graphene*

In the case of pure graphene our results show a perfect flat honeycomb structure without any ripples because of the relative small size of our computational unit cell. We found that C-atoms in the ground state have energy of 7.40 eV/atom with a lattice parameter of 2.46 Å and a C-C bond length of 1.42 Å (see Fig. 1). These lattice parameters are characteristic for $sp^2$ hybridization of carbon bonds in graphene and are in excellent agreement with experimental (for graphite 7.37 eV/atom [41], 7.4 eV/atom [42]) and computational parametric potential (7.40 eV/atom [31]) data.

Next we add atomic H and investigate the influence of these hydrogen atoms on the structure of graphene. Some high symmetric adsorption sites of a single hydrogen atom on graphene are shown in Fig. 2a: on top (T), twofold bridge (B), between top site and hollow (hexagonal center) (V), and between bridge and hollow (S).

The top position of the hydrogen atom changes the hybridization from $sp^2$ to $sp^3$ for the C atom bonded with H. This change of hybridization causes a local structural change in the graphene sheet (orange, yellow and red colored atoms shown in Fig. 2b). This is made more explicitly in Fig. 2c,d where we show a top and side view of the nearest atoms to the H-atom. The C atom (pink circle, Fig. 2d) bonded with hydrogen is shifted 0.78 Å above the plane of graphene, the C-C bond length with its three neighbors (orange circles) becomes 1.52 Å instead of 1.42 Å, and these three atoms are shifted by 0.33 Å above the plane of graphene. The next six neighbors (yellow circles) are shifted by 0.20 Å and the next three (red circles) by 0.17 Å above the plane of graphene.

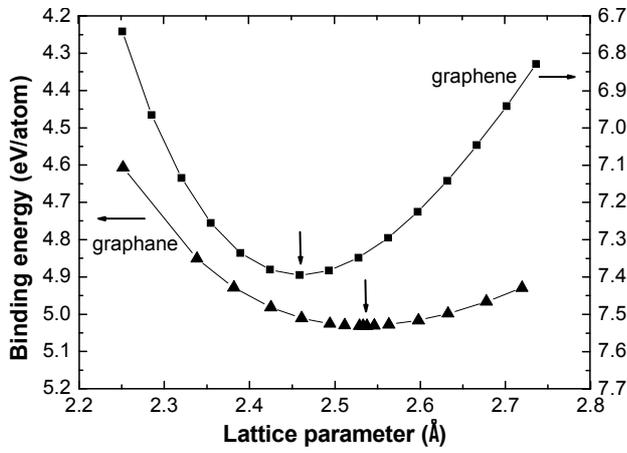

Fig. 1. Binding (cohesive) energy versus the lattice parameter of pure graphene and graphane. Vertical arrow indicates the lattice parameter corresponding to minimal energy of the system.

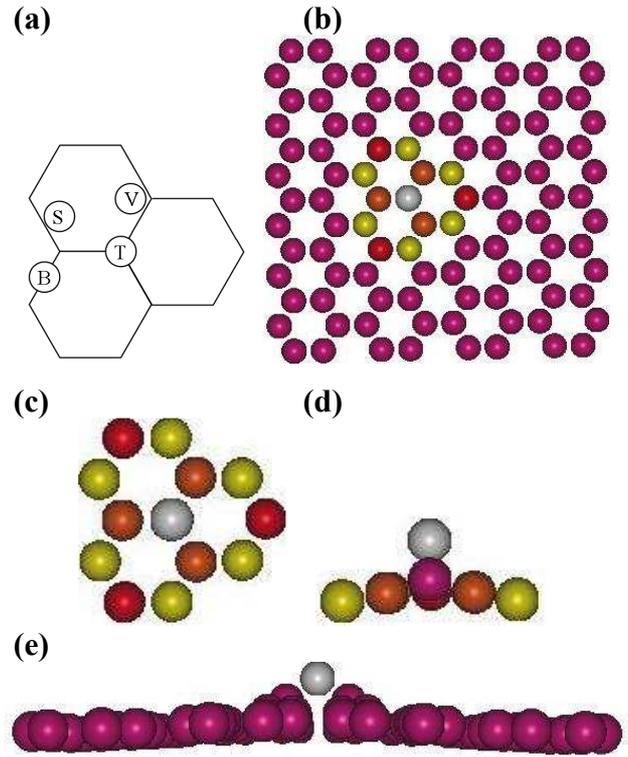

Fig. 2. The considered H atom positions on graphene (a): top site (T), bridge site (B), between top and hollow (hexagonal center) site (V), between bridge and hollow site (S). Top (b,c) and side (d) view of T, and side view of B (e) positions of H atom (grey color) after relaxation of the total system.

The C-H bond length is 1.09 Å which is characteristic for most CH systems, and the hydrogen atom is 1.87 Å above the graphene layer. The H-C-C bond angle is $107.2^0$ which is close to the one ($107.9^0$) calculated by us for a free $C_4H$ cluster and a little smaller than the bond angle for $sp^3$ hybridization ($109.5^0$) in the case of $CH_4$ and diamond (see Table 1). The C-C-C bond angle for the C atom beneath the H atom is $111.6^0$. This value is also typical for a $C_4H$ cluster. For this top site (T) case the binding energy of the H atom is 1.10 eV which is in good agreement with recent DFT data 0.98 eV [43] obtained by the SIESTA code. However, note that the binding energy for a single hydrogen atom reported in publications varies significantly (0.38-0.83 eV [44], 0.57 eV [45], 0.67 eV [11,22], 0.79 eV [5], 0.48 eV [23], 0.83 eV [24,46], 0.87 eV [15], 1.44 eV [16], 1.44-2.07 eV [47]) due to different methods and computational setups (in particular, simulation cell sizes) used in these studies and the same tendency is observed for H pairs [28,44]. Besides, the height of the C atom that is beneath the H atom obtained by DFT varies from 0.3 Å [16,23] (in most cases) to 0.52 Å [48].

In the case of a twofold bridge site (Fig. 2e) the H atom is 1.60 Å above the graphene layer, and due to its bonds with two carbon atoms the C-H bond length is increased a little: to 1.22 Å. Its two carbon neighbors are pushed 0.88 Å above the graphene plane, their distance is larger (1.98 Å), and the C-C bond length with their next four neighbors which are at a height of 0.50 Å – are a little shorter (1.397 Å) in comparison with bulk graphene 1.42 Å. In this case the binding energy of the H atom is -6.11 eV which implies an unbound metastable state. This adsorption site causes a distortion of the graphene sheet in a larger area around the H atom in comparison with the top position. The other two, i.e. cases V and S, are also metastable states with H binding energy -4.37 eV and -4.15 eV, respectively. In these cases the C-H bond lengths are 1.29 Å and 1.336 Å, the height of the adsorption sites are 0.94 Å and 0.86 Å, respectively. In earlier studies the V, S [22] and B [22,45] sites were also found to be metastable states for H adsorption.

In order to evaluate the influence of a second hydrogen atom on the graphene layer that contains already a single H atom we studied several top and twofold bridge sites of two hydrogen atoms as depicted in Fig. 3. Note that in the case of a single adsorbed atom, the hydrogen atom is exactly on top of a carbon atom. However, in the cases of the on top positions of the two hydrogen atoms, the C-H axis is tilted from the normal of the surface (see Fig. 3b) due to the bond angle of the $sp^3$ hybridization as both underlying C atoms are shifted by the same value above the graphene layer. The tilt angle is $12.8^0$ (or the H-C-C bond angle is $102.8^0$ to the direction of the next H atom), $5.1^0$ and $2.3^0$ for *ortho* (T1), *meta* (T2) and *para* (T3) hydrogen pairs, respectively. For T1 the C-C bond length of the underlying C atoms is 1.57 Å and the separation between the two H atoms is 2.05 Å which is close to the reported theoretical one (2.11 Å) [24]. In the case of T2 the separation between the H atoms is 2.77 Å (2.65 Å [24]) and the one for the underlying C atoms is 2.57 Å which compares to 2.46 Å for pure graphene.

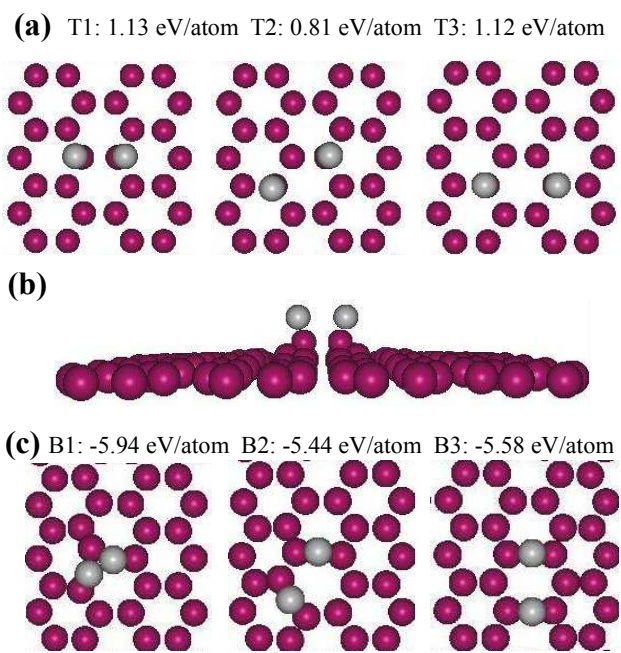

Fig. 3. The top (a, b) and bridge (c) sites for H pairs (grey color). The binding energy of the H-pair is shown on top of each figure.

These configurations are ordered as T1, T3, T2 with hydrogen binding energies 1.13, 1.12, 0.81 eV/atom, respectively which are in good agreement with earlier reported DFT data 1.19, 1.17 and 0.67 eV/atom [26]. The same order of energies for ortho, para and meta H attachments, but with slightly large binding energies of 1.47, 1.47 and 0.72 eV/atom were found in [5], 1.40, 1.37 and 0.82 eV/atom in [27] as well as 1.37, 1.36 and 0.82 eV/atom in [44] were found by other DFT calculations. Authors of Ref. [44] obtained also the hydrogen binding energies 1.05, 0.95 and 0.42 eV/atom for T3, T1 and T2, respectively in their earlier work [46] which can be explained due to the different computational setups used in these two studies. So, the most stable configuration for hydrogen dimer (T1) predicted in our calculation is in a good agreement with other theoretical data and experimentally found *ortho* hydrogen pairs [25], although there is no big energy difference between *ortho* (T1) and *para* (T3) hydrogen pairs. The weaker binding energy in the *meta* configuration as compared to the *ortho* or *para* configurations is explained by the bonding of both H atoms to the same subgroup of C atoms, which creates two radicals, i.e. unpaired electrons [5].

For two absorbed H atoms we may consider another situation where the two H atoms are chemisorbed on each side of the graphene layer. In this case our calculation shows that the most energetically favorable structure corresponds to an *ortho* hydrogen pair with much higher binding energy (1.27 eV/atom) than the *para* hydrogen pair (1.07 eV/atom). This tendency on energy ordering for adsorption of two H on the same side and on the both sides of graphene agrees very well with results presented in [5]. The case of H chemisorption on both sides of graphene studied in detail in [5] will be further discussed later.

These results show that the local structure of the graphene layer becomes more distorted when both hydrogen atoms are close to each other, which in itself is not surprising. The same picture is observed for the twofold bridge sites: the most distorted is for case B1. The height of the H sites is 1.93 Å, 1.86 Å, and 1.75 Å for B1, B2, and B3, respectively. The C-H bond lengths are 1.24 Å and 1.27 Å with the carbon atoms on both sides of H for B1, 1.15 Å and 1.30 Å for B2. It is 1.22 Å for B3 as in the case of a single H on the bridge site, but for B3 the structure of the graphene layer is more distorted especially in the direction of the H-H axis. As the hydrogen binding energy is negative for the bridge (B1-B3) sites, they are unbound.

As bridge adsorption sites are unfavorable, further we will limit ourselves to the top position of adsorbed hydrogen atoms. In Fig. 4 three hydrogen atoms adsorbed on top position within one benzene ring (T1-T3) as well as on two or three carbon hexagons (T4-T6) are shown. Following our calculations the most energetically favorable configuration in these cases is also *ortho* hydrogen pairs, i.e. the most H packed structure T1 with the H binding energy 1.15 eV/atom. The tilt angle of the mid H atom is rather large: $34.8^0$, for the other two H atoms it is only ~$16^0$. The H-C-C bond angle for this atom in its tilting direction is $81.3^0$, although in the other directions it is close to the one of a *sp³* hybridization. The next stable configuration with almost the same H binding energy (1.14 eV/atom) is an *ortho-para* combination (T6). Then the *para* pair configuration T4 of the H trimer has a slightly lower binding energy of 1.12 eV/atom. The next T5 structure has a hydrogen binding energy 1.06 eV/atom and corresponds to S5 [27] with 1.31 eV/atom which was suggested as an explanation of the observed starlike shape of STM images [25]. However, it is questionable that *meta* configurations labeled as S4 and S6 in [27] have almost the same energy as S5. In all our considered cases this *meta* pair configuration T3 is the least stable which agrees with [49]. The results show that not only a pure *meta* pair configuration, but even any its combination with *ortho* or *para* pairs, is less stable as indicated in T2 of Fig. 4.

An interesting case is the one when the number of H atoms is equal to 6, i.e. equal to the number of C atoms in one hexagon. Some possible lattice geometries of these H atoms on top sites are shown in Fig. 5. Possible configurations consist of hexagons of hydrogen atoms oriented in phase with the carbon hexagon (see cases T1 and T2, Fig. 5a,b).

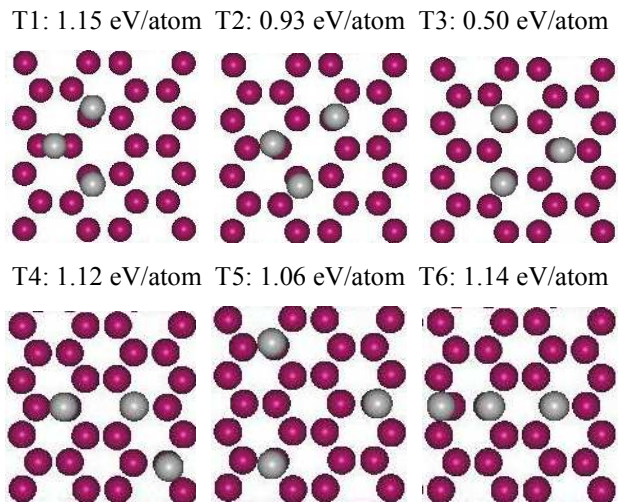

Fig. 4. Some top sites for three H atoms. The H binding energy per atom is shown on top of each figure.

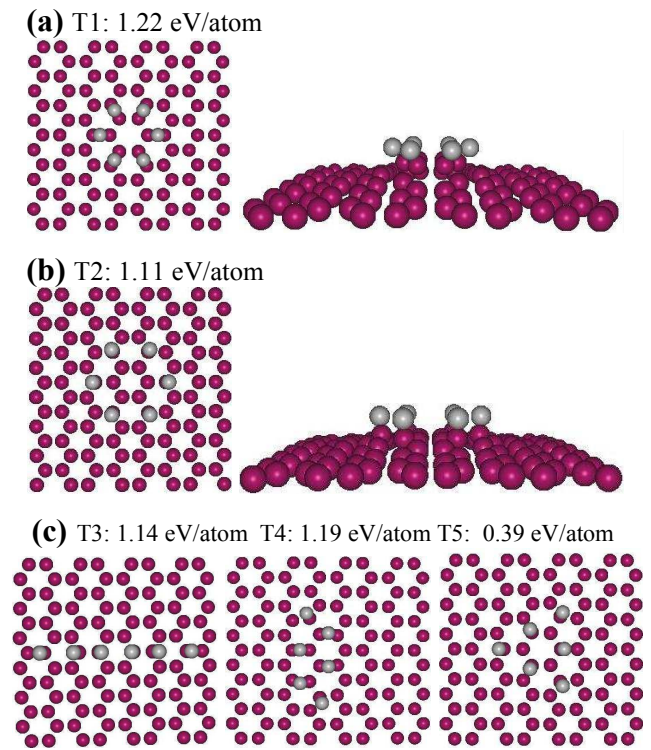

Fig. 5. Some top site structures for six H atoms: small (a) and big (b) hexagonal; linear, zigzag and trigonal (c). On top of figures the H binding energy per atom.

The C-C bond length in the T1 configuration is 1.51 Å which is considerable larger than 1.42 Å in pristine graphene. The C hexagon is pushed 1.34 Å out of the graphene layer, the C-C bond length for its next six carbon neighbors is 1.59 Å which are shifted 0.54 Å from the surface. The H-hexagon is 2.20 Å above the graphene layer which is rather large: the separation between the nearest neighbor H atoms is 2.22 Å. However, they are only bonded with C atoms with a C-H bond length of 1.11 Å, with a H-C-C bond angle of $80.8^0$ and $108.4^0$ in their tilted and two other directions, respectively. As indicated in Fig. 5 this hydrogen hexagon T1 formed by *ortho* pairs is the most stable configuration of six H atoms adsorbed on one side of the graphene sheet.

The next stable H adsorption structure is an *ortho* pair zigzag configuration T4 followed by the *ortho-para* pair structure T3. In this linear structure T3 the graphene layer is bent upward along the linear hydrogen chain. The carbon chain bonded with the hydrogen atoms is shifted significantly above the graphene layer pulling two to three next carbon atom rows out of the graphene plane. A similar structure will be discussed in more detail in the next subsection. A *para* pair hexagon structure depicted as T2 in Fig. 5b has the H binding energy of 1.11 eV/atom which is 0.11 eV/atom lower than for T1. As mentioned for H dimer and trimer, a *meta* pair structure T5 (Fig. 5c) is energetically the least favorable. This configuration is given for completeness.

As shown for a H dimer, the largest hydrogen binding energy (1.33 eV/atom) in this case also corresponds to the chemisorption structure of *ortho* hydrogen pairs adsorbed on both sides of graphene within one hexagon. In this case the structure is buckled within this hexagon which pushes the hydrogen-connected carbon atoms out (above and below) of the graphene plane. As shown in [5], this configuration is the most stable of all the above considered ones due to the relaxation of the lattice around the H atoms.

*3.2 From linear hydrogenation to graphane*

Now we consider linear chains of H atoms bound to the graphene layer. There are two possible positions: along a zigzag direction (Fig. 6a,c) and along an armchair direction (Fig. 6b,d). For the zigzag direction (Fig. 6a) we allow for relaxation of the separation between the H-atoms, but find that both initial and final separations are identical: 2.46 Å and that the H-chain stays straight. This structure with H binding energy of 0.45 eV/atom is energetically less favorable as it is formed by hydrogen *meta* pairs. The graphene layer is bent upward along this linear structure as the H atoms form $sp^3$ hybridization affecting the closest carbon atom row and its neighbor rows.

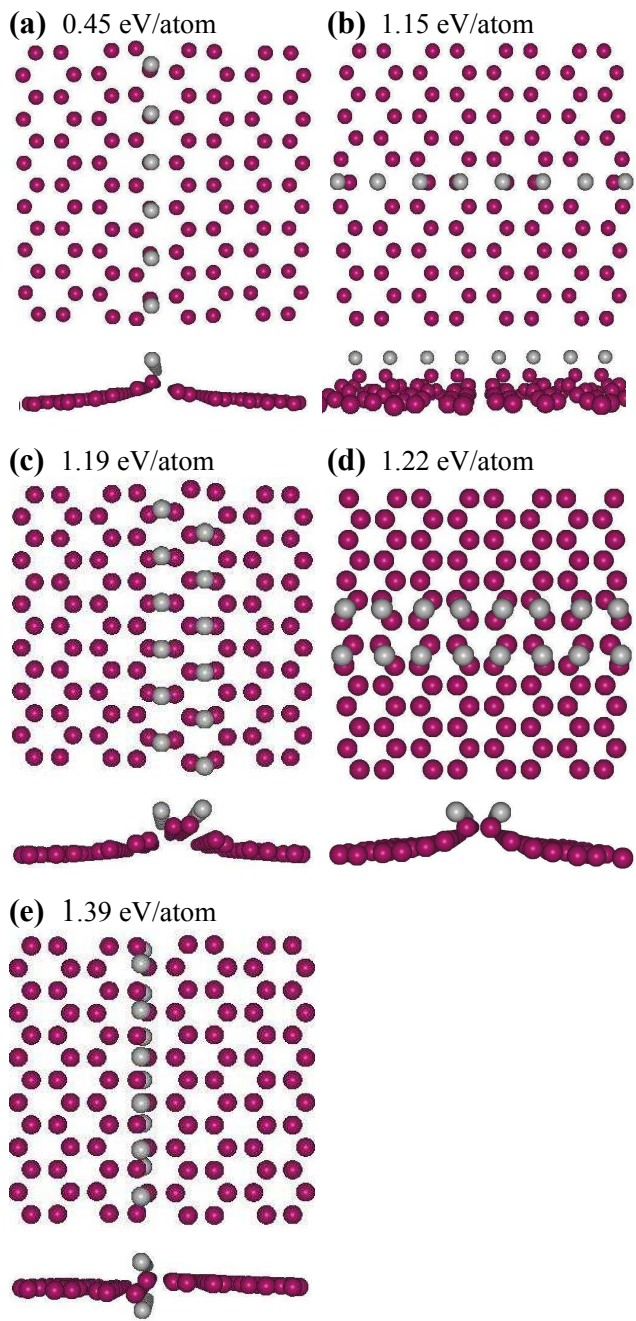

Fig. 6. Top (top) and side (bottom) views of linear H structures on graphene along zigzag (a, c) and armchair (b, d) directions, and for H chains on both sides (e). On top of figures the H binding energy per atom.

Table 2. The hydrogen binding energy for the most preferential adsorption sites for different number of adsorbed hydrogen.

|  | H-binding energy (eV/atom) |
|---|---|
| Initial H adsorption stages: |  |
| single H (T) | 1.10 |
| double H (T1) | 1.13 |
| triple H (T1) | 1.15 |
| six H (T1) | 1.22 |
| double *ortho* H on both sides | 1.27 |
| six *ortho* H on both sides | 1.33 |
| Regular H adsorption structures: |  |
| linear *ortho-para* H chain | 1.15 |
| double *ortho-para* H chain | 1.22 |
| linear H chains on both sides | 1.39 |
| single-sided half coverage (boat) | 0.52 |
| single-sided full coverage | 0.34 |
| Graphane | 1.50 |

For the armchair case the initial separations between two H (or C) atoms along the row are 1.42 Å and 2.84 Å (Fig. 6b). In this case due to the linear H chain the separation between C atoms changes by 0.13 Å after relaxation: now they are 1.55 Å and 2.71 Å. Meanwhile the separations between H atoms are close to each other after relaxation: 2.05 Å and 2.21 Å, i.e. the unit cell of the H-chain consists of two H-atoms. This structure has a higher H binding energy (1.15 eV/atom) than the one for the zigzag case (Fig. 6a).

Next we consider a pair of H chains. Along the zigzag and armchair directions, a pair of parallel H chains initially separated (i.e. in its unrelaxed state) by distances of 0.71 Å and 1.23 Å were studied (Fig. 6c,d). In both structures the H binding energy is almost the same. As mentioned above although a single H chain along a zigzag direction is energetically less favorable, this pair of H chains (Fig. 6c) formed by *ortho* pairs is more favorable. In the case of 0.71 Å initial separation the distance between the two rows of carbon atoms beneath the H chains increases only by 0.13 Å, i.e. it becomes 0.84 Å. However, the distance between the hydrogen rows becomes 2.24 Å. The T1 structure of three H atoms (see Fig. 4) and the T4 structure of six H atoms (see Fig. 5) are the unit cell of this zigzag structure. Along the zigzag direction, increasing the separation between the H chains further results in an energetically unfavorable hydrogenated structure. In case of two H-chains along the armchair direction the distance between the two rows of carbon atoms beneath the H chains increases by 0.15 Å, i.e. it becomes 1.38 Å. However, the distance between the hydrogen rows becomes too large, i.e. 2.54 Å. We found that both, the two carbon and two hydrogen atom rows are shifted above the graphene layer. Note that as shown in [50] hydrogen atoms tend to form one-dimensional chains with ripples made up of *sp²* carbon atoms between them and it is possible to turn graphene into a semiconductor by changing the separation between the H-chains [51].

All the above considered cases for hydrogenation of a single graphene layer on one side. Now we turn to the case when hydrogenation occurs on *both sides of the graphene layer* as shown in Fig. 6e. In this case we find displacements of several C rows above and below the graphene layer. The C-H axis is tilted by $16.1^0$ from the normal of the graphene plane. This structure has a higher binding energy (1.39 eV/atom) as compared to the previous structures of a single or two H rows on the same side of the graphene layer (Table 2). The tilt of the C-H axis disappears when one more H row is added next to the existing H rows. In this case each C row beneath the H atoms is shifted towards the H rows and the graphene layer becomes buckled.

Such buckled structure was also found in cluster expansion calculations on H chemisorbed islands onto graphene [12] and in recently predicted chair-like conformation of graphane [13]. We obtained a similar chair conformer of graphane shown in Fig. 7. This buckled structure is caused by the transition from the planar $sp^2$ hybridization in the case of pure graphene to the tetrahedral $sp^3$ hybridization provided by H atoms in the case of graphane. The obtained C-C bond length of 1.54 Å is in good agreement with the one reported in [13,16] and corresponds to the one for $sp^3$ hybridization of C. The energy per atom of this conformer versus the lattice parameter of the graphene matrix is shown in Fig. 1. The cohesive energy of graphane in the ground state is 5.03 eV/atom and results in the highest H binding energy which is 1.50 eV/atom. The lattice parameter of 2.53 Å is 3% larger than the one for pure graphene and much larger than 2.42 Å reported in [29].

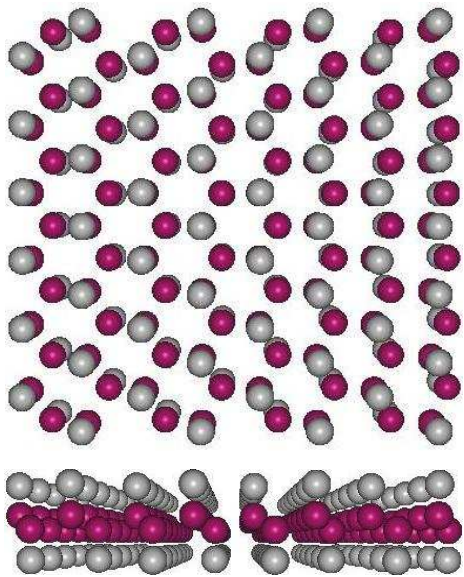

Fig. 7. Top (top) and side (bottom) views of chair-like conformer of graphane (H atoms - grey).

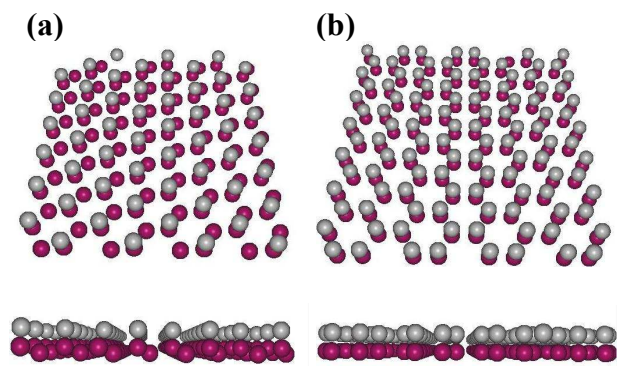

Fig. 8. Half (a) and full (b) hydrogen (grey) coverage on one side of the graphene layer.

Next we studied the cases of single-side half and full hydrogenation of a graphene layer. In the case of half coverage the hydrogen atoms are bonded with the carbon atoms similar to the case of graphane, but only on one side as shown in Fig. 8a and referred to as "*graphone*" in [52]. In this case the carbon atoms bonded with H are displaced above the carbon atoms which have no H neighbors. The C-C bond length and the lattice parameter are almost the same as in graphane: 1.53 Å and 2.58 Å, respectively. However, the one side half H coveraged graphene structure when considered as a *meta* pair configuration has a H binding energy of -0.57 eV/atom which is unstable. However, a boat-like conformer of graphane coveraged by *ortho-para* pair H chains on only one side (not shown) has a H binding energy 0.52 eV/atom. As shown in [52], *graphone* is a ferromagnetic semiconductor with a small indirect gap. In the case of full coverage on one side, i.e. when all carbon atoms are bonded with a single H atom on top position (Fig. 8b) both the C-C bond length and the lattice parameter (1.65 Å and 2.87 Å, respectively) are much larger than the ones for pure graphene which causes some doubt about the existence of such a structure. There is no buckling in this structure, and the C-H bond length is 1.12 Å. This structure has a H binding energy 0.34 eV/atom which is less than the one for graphane. Although such a single-sided structure was claimed to be found in a recent experiment [29] (however, note that there was not any information about the hydrogen concentration in that paper), the earlier theoretical calculations [13,16] found that they were unstable. This clearly calls for further theoretical and experimental investigations.

## 4. Conclusions

Atomistic simulations were used to optimize the structure and energetics of hydrogen chemisorbed on a single graphene layer. The most stable chemisorption sites of H atoms and the structural changes of a hydrogenated graphene layer were found for one, two, three and six H atoms. We found that H atoms are preferentially adsorbed on the top site of the C atoms for a single H atom, and an *ortho* hydrogen pair is formed for two and more H atoms. We also studied chains of H atoms adsorbed on graphene as well as half and full saturated hydrogenation. We found that a single-sided linear H chain along a zigzag direction is energetically less favorable as it is formed by hydrogen *meta* pairs, although a single-sided two neighbor H chains can exist. These results show that single-sided hydrogenation is possible for low hydrogen concentration. Our results are summarized in Table 2. It is seen that in all cases a double-sided H structure has the highest binding energy. We found that single-sided half coverage *meta* pair hydrogenation is unstable, while a hydrogenation by *ortho-para* pair H chains can exist. The most stable adsorbed structures are ordered as *ortho* pair, *orth-para* pair combination and *para* pair configurations. It means that in case of random hydrogenation various H structures can be simultaneously observed in the same sample (as shown in [53]) each occurring with a probability determined by their energy.

The structure of the graphene sheet used in the present study and the regular hydrogenated structures presented in Sec. 3.2 are systems of infinite size in the lateral plane. Hydrogenation of a free-standing graphene sheet of finite size is different from the present study and will be subject of future investigation. Moreover, as chemisorbed H atoms causes besides structural alterations also changes of the electrophysical properties of the graphene layer its engineering by hydrogenation may be used to modify graphene on a nanoscale.

**Acknowledgment.** AD thanks M.W. Zhao for a useful correspondence. This work was supported by the Belgian Science Policy (IAP) and the Flemish Science Foundation (FWO-Vl).

**References**

[1] Novoselov KS, Geim AK, Morozov SV, Jiang D, Zhang Y, Dubonos SV, Grigorieva IV, Firsov AA. Electric field effect in atomically thin carbon films. Science 2004;306:666-9.

[2] Dillon AC, Jones KM, Bekkedahl TA, Kiang CH, Bethune DS, Heben MJ. Storage of hydrogen in single-walled carbon nanotubes. Nature 1997;386:377-9.

[3] Dillon AC, Heben MJ. Hydrogen storage using carbon adsorbents: past, present and future. Appl Phys A 2001;72:133-42.

[4] Hirscher M, Becher M, Haluska M, Dettlaff-Weglikowska U, Quintel A, Duesberg GS, et al. Hydrogen storage in sonicated carbon materials. Appl Phys A 2001;72:129-32.

[5] Lin Y, Ding F, Yakobson BI. Hydrogen storage by spillover on graphene as a phase nucleation process. Phys Rev B 2008;78:041402(R)-1-4.

[6] Hynek S, Fuller W, Bentley J. Hydrogen storage by carbon sorption. Int J Hydrogen Energy 1997;22:601-10.

[7] Liu C, Fan YY, Liu M, Cong HT, Cheng HM, Dresselhaus MS. Hydrogen storage in single-walled carbon nanotubes at room temperature. Science 1999;286:1127-9.

[8] Lee SM, Lee YH. Hydrogen storage in single-walled carbon nanotubes. Appl Phys Lett 2000;76:2877-9.

[9] Bünger U, Zittel W. Hydrogen storage in carbon nanostructures – still a long road from science to commerce? Appl Phys A 2001;72:147-51.

[10] Ruffieux P, Gröning O, Bielmann M, Mauron P, Schlapbach L, Gröning P. Hydrogen adsorption on $sp^2$-bonded carbon: Influence of the local curvature. Phys Rev B 2002;66:245416-1-8.

[11] Sha X, Jackson B. First-principles study of the structural and energetic properties of H atoms on a graphite (0001) surface. Surf Sci 2002;496:318-30.

[12] Sluiter MHF, Kawazoe Y. Cluster expansion method for adsorption: Application to hydrogen chemisorption on graphene. Phys Rev B 2003;68:085410-1-7.

[13] Sofo JO, Chaudhari AS, Barber GD. Graphane: A two-dimensional hydrocarbon. Phys Rev B 2007;75:153401-1-4.

[14] Yazyev OV, Helm L. Defect-induced magnetism in graphene. Phys Rev B 2007;75:125408-1-5.

[15] Lehtinen PO, Foster AS, Ma Y, Krasheninnikov AV, Nieminen RM. Irradiation-induced magnetism in graphite: A density functional study. Phys Rev Lett 2004;93:187202-1-4.

[16] Boukhvalov DW, Katsnelson MI, Lichtenstein AI. Hydrogen on graphene: Electronic structure, total energy, structural distortions and magnetism from first-principles calculations. Phys Rev B 2008;77:035427-1-7.

[17] Yu D, Liu F. Synthesis of carbon nanotubes by rolling up patterned graphene nanoribbons using selective atomic adsorption. Nano Lett 2007;7:3046-50.

[18] Singh AK, Yakobson BI. Electronics and magnetism of patterned graphene nanoroads. Nano Lett 2009;9:1540-43.

[19] Singh AK, Penev ES, Yakobson BI. Vacancy clusters in graphane as quantum dots. ACS Nano 2010;4:3510-14.

[20] Ribas MA, Singh AK, Sorokin PB, Yakobson BI. Patterning nanoroads and quantum dots on fluorinated graphene. Nano Res 2011;4:143-52.

[21] Ryu S, Han MY, Maultzsch J, Heinz TF, Kim P, Steigerwald ML, Brus LE. Reversible basal plane hydrogenation of graphene. Nano Lett 2008;8:4597-602.

[22] Miura Y, Kasai H, Dino WA, Nakanishi H, Sugimoto T. Effective pathway for hydrogen atom adsorption on graphene. J Phys Soc Jpn 2003;72:995-7.

[23] Ishii A, Yamamoto M, Asano H, Fujiwara K. DFT calculation for adatom adsorption on graphene sheet as a prototype of carbon nano tube functionalization. J Phys: Conference Series 2008;100:052087-1-4.

[24] Roman T, Dino WA, Nakanishi H, Kasai H, Nobuhara K, Sugimoto T, Tange K. Identifying hydrogen atoms on graphite. J Phys Soc Jpn 2007;76:114703-1-4.

[25] Hornekær L, Xu W, Otero R, Lægsgaard E, Besenbacher F. Long range orientation of metastable atomic hydrogen adsorbate clusters on the graphite(0001) surface. Chem Phys Lett 2007;446:237-42.

[26] Roman T, Dino WA, Nakanishi H, Kasai H, Nobuhara K, Sugimoto T, Tange K. Hydrogen pairing on graphene. Carbon 2007;45:218-20.

[27] Khazaei M, Bahramy MS, Ranjbar A, Mizuseki H, Kawazoe Y. Geometrical indications of adsorbed hydrogen atoms on graphite producing star and ellipsoidal like features in scanning tunneling microscopy images: Ab initio study. Carbon 2009;47:3306-12.

[28] de Andres PL, Vergés JA. First-principles calculation of the effect of stress on the chemical activity of graphene. Appl Phys Lett 2008;93:171915-1-3.

[29] Elias DC, Nair RR, Mohiuddin TMG, Morozov SV, Blake P, Halsall MP, et al. Control of graphene's properties by reversible hydrogenation: Evidence for graphane. Science 2009;323:610-13.

[30] Ray NR, Srivastava AK, Grotzschel R. In search of graphane - A two-dimensional hydrocarbon. *arXiv*: Cond Mat Mtrl Sci 0802.3998v1; 2008.

[31] Tewary VK, Yang B. Parametric interatomic potential for graphene. Phys Rev B 2009;79:075442-1-9.

[32] Brenner DW, Shenderova OA, Harrison JA, Stuart SJ, Ni B, Sinnott SB. A second-generation reactive empirical bond order (REBO) potential energy expression for hydrocarbons. J Phys: Condens Matter 2002;14:783-802.

[33] Kosimov DP, Dzhurakhalov AA, Peeters FM. Theoretical study of the stable states of small carbon clusters $C_n$ (*n*=2–10). Phys Rev B 2008;78:235433-1-8.

[34] Yamayose Y, Kinoshita Y, Doi Y, Nakatani A, Kitamura T. Excitation of intrinsic localized modes in a graphene sheet. Europhys Lett 2007;80:40008-1-6.

[35] Lu Q, Arroyo M, Huang R. Elastic bending modulus of monolayer graphene. J Phys D: Appl Phys 2009;42:102002-1-6.

[36] Reddy CD, Rajendran S, Liew KM. Equilibrium configuration and continuum elastic properties of finite sized graphene. Nanotechnology 2006;17:864-70.

[37] Maslov MM, Podlivaev AI, Openov LA. Nonorthogonal tight-binding model for hydrocarbons. Phys Lett A 2009;373:1653-7.

[38] Allen PW, Sutton LE. Tables of interatomic distances and molecular configurations obtained by electron diffraction in the gas phase. Acta Cryst 1950;3:46-72.

[39] van Nes GJH, Vos A. Single-crystal structures and electron density distributions of ethane, ethylene and acetylene. III.* Single-crystal X-ray structure determination of ethylene at 85 K. Acta Cryst B 1979;35:2593-601.

[40] Feldman T, Romanko J, Welsh HL. The $v_2$ Raman band of methane. Can J Phys 1955;33:138-45.

[41] Kittel C. Introduction to Solid State Physics. 5th ed. New York: Wiley; 1976.

[42] Ebbesen TW. Carbon nanotubes: Preparation and properties. New York: CRC press; 1997.

[43] Miwa RH, Martins TB, Fazzio A. Hydrogen adsorption on boron doped graphene: an *ab initio* study. Nanotechnology 2008;19:155708-1-7.

[44] Šljivančanin Ž, Rauls E, Hornekær L, Xu W, Besenbacher F, Hammer B. Extended atomic hydrogen dimer configurations on the graphite (0001) surface. J Chem Phys 2009;131:084706-1-6.

[45] Jeloaica L, Sidis V. DFT investigation of the adsorption of atomic hydrogen on a cluster-model graphite surface. Chem Phys Lett 1999;300:157-62.

[46] Hornekær L, Rauls E, Xu W, Šljivančanin Ž, Otero R, Stensgaard I, Lægsgaard E, Hammer B, Besenbacher F. Clustering of chemisorbed H(D) atoms on the graphite (0001) surface due to preferential sticking. Phys Rev Lett 2006;97:186102-1-4.

[47] Li WF, Zhao MW, He T, Song C, Lin XH, Liu XD, Xia YY, Mei LM. Concentration dependent magnetism induced by hydrogen adsorption on graphene and single walled carbon nanotubes. J Magn Magn Mater 2010;322:838-43.

[48] Wessely O, Katsnelson MI, Nilsson A, Nikitin A, Ogasawara H, Odelius M, Sanyal B, Eriksson O. Dynamical core-hole screening in the x-ray absorption spectra of hydrogenated carbon nanotubes and graphene. Phys Rev B 2007;76:161402(R)-1-4.

[49] Roman T, Nakanishi H, Kasai H, Nobuhara K, Sugimoto T, Tange K. Stability of three-hydrogen clusters on graphene. J Phys Soc Jpn 2009;78:035002-1-2.

[50] Xiang HJ, Kan EJ, Wei S-H, Gong XG, Whangbo M-H. Thermodynamically stable single-side hydrogenated graphene. Phys Rev B 2010;82:165425-1-4.

[51] Chernozatonskii LA, Sorokin PB, Brüning JW. Two-dimensional semiconducting nanostructures based on single graphene sheets with lines of adsorbed hydrogen atoms. Appl Phys Lett 2007;91:183103-1-3.

[52] Zhou J, Wang Q, Sun Q, Chen XS, Kawazoe Y, Jena P. Ferromagnetism in semihydrogenated graphene sheet. Nano Lett 2009;9:3867-70.

[53] Balog R, Jørgensen B, Wells J, Lægsgaard E, Hofmann P, Besenbacher F, Hornekær L. Atomic hydrogen adsorbate structures on graphene. J Am Chem Soc 2009;131:8744-45.